\documentclass[twocolumn,showpacs,superscriptaddress,nofootinbib]{revtex4-1}
\usepackage[T1]{fontenc} % if needed
\usepackage{amsmath}
\usepackage{amssymb}
\usepackage{amsfonts}
\usepackage{graphicx}
\usepackage{enumitem}
\usepackage{bm}
\usepackage{rotating}
\usepackage{hyperref}
\usepackage{xcolor}
\usepackage{array}
\usepackage{lmodern}
\usepackage[normalem]{ulem}
\usepackage{braket}
\usepackage{tensor}
\usepackage{mathrsfs}
\usepackage{float}

\newcommand{\be}{\begin{equation}}
\newcommand{\ee}{\end{equation}}
\newcommand{\ba}{\begin{eqnarray}}
\newcommand{\ea}{\end{eqnarray}}
\newcommand{\beq}{\begin{equation}}
\newcommand{\eeq}{\end{equation}}
\newcommand{\beqa}{\begin{eqnarray}}
\newcommand{\eeqa}{\end{eqnarray}}

\newcommand{\sx}{\mathsf{x}}

\usepackage[normalem]{ulem}

\begin{document}

\title{Quantum Detection of  Conicity}

\author{Wan Cong}
\email{wcong@uwaterloo.ca}
\address{Department of Physics and Astronomy, University of Waterloo, Waterloo,
Ontario, Canada, N2L 3G1}
\address{Perimeter Institute, 31 Caroline St., Waterloo, Ontario, N2L 2Y5, Canada}

\author{Ji\v r\'i Bi\v c\'ak}
\email{bicak@mbox.troja.mff.cuni.cz }
\address{Institute of Theoretical Physics, Faculty of Mathematics and Physics, Charles University, V Hole\v sovi\v ck\'ach 2, 180 00 Prague 8, Czech Republic}

\author{David Kubiz\v n\'ak}
\email{dkubiznak@perimeterinstitute.ca}
\address{Perimeter Institute, 31 Caroline St., Waterloo, Ontario, N2L 2Y5, Canada}
\address{Department of Physics and Astronomy, University of Waterloo, Waterloo,
Ontario, Canada, N2L 3G1}

\author{Robert B. Mann}
\email{rbmann@uwaterloo.ca}
\address{Department of Physics and Astronomy, University of Waterloo, Waterloo,
Ontario, Canada, N2L 3G1}
\address{Perimeter Institute, 31 Caroline St., Waterloo, Ontario, N2L 2Y5, Canada}

\begin{abstract}
We investigate the sensitivity of an Unruh--DeWitt detector to the global features of a deficit angle that are otherwise classically inaccessible.
Specifically,  we consider a detector placed inside an infinite thin hollow cylinder whose spacetime is everywhere flat but outside of which the spacetime has a
  deficit angle and study its response to a scalar field to which it couples. We find that the response of the detector is sensitive to the deficit angle, despite the fact that it does not interact with the cylinder.

\end{abstract}

\maketitle

\section{Introduction}

The Unruh--DeWitt (UDW) detector \cite{Unruh:1976db,DeWitt:1980hx} was first introduced to give an operational meaning to acceleration radiation (in the context of the Unruh effect) and to particle detection in curved spacetimes. Today, we can use it to probe (theoretically) various global features of the spacetime. For example, the response of a detector placed in different quotient spaces of Minkowski space has been shown to be sensitive to spacetime topology \cite{Langlois:2005nf,PhysRevD.93.044001}, and their sensitivity to topological features hidden behind event horizons has likewise been demonstrated   \cite{Smith:2013zqa,Ng:2017iqh}. UDW detectors have also been shown to be able to detect the presence of a massive spherical shell enveloping the   inertial detectors 
\cite{Ng:2016hzn, cong:2020crf} and the corresponding frame dragging in the case of a rotating shell \cite{Cong:2020xlt} in situations not possible for classical measuring devices.

 In the present paper we focus on   quantum detection of spacetime conicity.
Specifically, we are interested in studying the sensitivity of a UDW detector to the global features of a deficit angle when its quasilocal manifestation is absent. We do this by placing the detector inside an infinite thin hollow cylinder (cylindrical shell), described by a special case of the Levi-Civita metric \cite{LC}, whose 
spacetime is flat everywhere, but has a deficit angle \textit{outside} the cylinder. Being induced by the energy-momentum of the cylinder, the deficit angle is \textit{not present inside} the cylinder (where the detector is situated) and the axis is regular. This is in contrast to the usual idealized (distributional) cosmic string spacetime \cite{Hindmarsh:1994re,Vilenkin:2000jqa}, where the conical deficit is present throughout\footnote{ The distributional cosmic string spacetime \cite{Vilenkin:1981zs} 
is recovered upon the limit of the vanishing radius of the cylinder. More elaborate models of cosmic strings where the distributional character is smoothed out include, for example, constant density models \cite{1985ApJ...288..422G},
the abelian Higgs model \cite{Nielsen:1973cs,PhysRevD.32.1323}, or the recent model \cite{Boos:2020kgj} which takes into account a non-local description of the gravitomagnetism.}.  The absence of a deficit angle inside the cylinder provides an interesting set-up to study the response of a UDW detector since an observer with access only to classical measuring devices in a finite-sized quasilocal region within the  cylinder could not detect its presence.

The effects of a ``global conical deficit'' on the vacua of quantum fields have been studied in the literature, such as the vacuum polarisation of the field \cite{Helliwell:1986hs,Frolov:1987dz,Linet:1987vz} and particle creation in these spacetimes \cite{Harari:1990xf,Skarzhinsky:1993gc}. UDW detectors and other quantum particles in cosmic string spacetimes have also been studied in the literature and it has been demonstrated that the detectors or particles are in general sensitive to the presence of a cosmic string \cite{Davies:1987th,PhysRevD.93.084028,AMIRKHANJAN1995197,Bilge:1997fn,Zhou:2018gqf,DeA.Marques:2007zz,Yang:2020bff,He:2020xhz}.
 Contrary to all these studies, in our setup the UDW detector is not situated in the region with the conical deficit; such a deficit is only present \textit{outside} the cylinder. 
  
% The formation of cosmic strings is predicted by some unified theories of particle interactions and their presence results in various interesting gravitational phenomena \cite{Zeldovich:1980gh,Vilenkin:1984ib,Gott:1984ef,Brandenberger:1993hw,PhysRevLett.85.3761,BHATTACHARJEE2000109}, making them an attractive area of active research. 
%

 The outline of our paper is as follows: Section \ref{sec: metric} details  the special form of the Levi-Civita metric we are using, Section \ref{sec: field} presents the quantisation of the massless scalar field, Section \ref{sec: UDW} computes the UDW detector response,  with the results displayed in Section \ref{sec: results}, and conclusions in Section \ref{sec: conclusion}. The details of the stress-energy tensor of the cylindrical shell are presented in Appendix \ref{appendix}.

\section{Cylinder spacetime}
\label{sec: metric}
Just as spherically symmetric massive shells are sources for the Schwarzchild spacetime, cylinders are sources for the Levi-Civita metric \cite{LC}. However, constructing physical cylinders and relating their properties (such as  mass) to the parameters of the Levi-Civita metric is more involved than in the spherical case, see, e.g., \cite{Bic_aacute_k_2002,Philbin:1995iz,Bicak:2004fw}. The metric we shall employ is the $M=0$ case of one studied by Bi\v c\'ak and \v Zofka \cite{Bic_aacute_k_2002}, which is globally flat. Inside the cylinder we have the usual Minkowski spacetime, with the metric written in cylindrical coordinates as:
\begin{equation}
\label{eq:in}
    ds^2 = -dt^2+dz^2+d\rho_-^2+\rho_-^2 d \phi^2\,,\quad 0\leq \rho_-\leq R_1,
\end{equation}
where $R_1$ is the proper radius of the cylinder. This metric will be matched on the cylinder to a special case of the Levi-Civita metric outside the cylinder, which is given by
\begin{equation}
\label{eq:out}
    ds^2 = -dt^2+dz^2+d\rho_+^2+\frac{\rho_+^2}{c^2} d \phi^2\,,\quad c R_1\leq \rho_+<c R_2,
\end{equation}
where $\phi\in[0,2\pi]$. In these coordinates, the cylinder is located at $\rho_+=c R_1$, which  ensures that the metrics 
\eqref{eq:in} and \eqref{eq:out} when induced on the cylinder agree, obeying the first Israel junction condition \cite{Israel1966}.
The second junction condition relates the stress-energy tensor of the cylinder to the jump in the extrinsic curvature across the cylinder. This stress-energy tensor has been worked out in \cite{Bic_aacute_k_2002}, where a more general metric for the spacetime outside the cylinder was used. We review the main results in Appendix \ref{appendix}. %\tcm{We also note that the metric above has been scaled by an arbitrary length factor $r_0$ (see Appendix \ref{appendix}), making the coordinates dimensionless. } 
The parameter $c$ in Eq.~\eqref{eq:out} describes the ``conicity'', see, e.g., \cite{Bic_aacute_k_2002,Bicak:2004fw} of the spacetime, and is related to a non-zero mass per unit length of the cylinder
\be\label{mu}
\mu=\frac{1}{4}\Bigl(1-\frac{1}{c}\Bigr)\,. 
\ee
In this paper, we are interested in the case $c>1$. Specifically, we can define a new angular coordinate $\varphi = \phi/c$, for which the metric in~\eqref{eq:out} reduces to the usual Minkowski metric in cylindrical coordinates but now $\varphi \in [0, 2\pi/c]$. In other words, the spacetime has a conical deficit of $\delta = 2\pi(1-1/c) >0$. 

In order to avoid dealing with asymptotics of infinite cylindrical systems, in what follows we will impose Dirichlet boundary conditions for the field on the surface $\rho_+ = c R_2$. This can be interpreted as a second infinitely long, perfectly reflecting cylinder, concentric to the first and having a proper radius of $R_2$ (see also \cite{PhysRevD.53.4382}). We will not concern ourselves with the metric outside this second cylinder as the Dirichlet boundary condition ensures that the field will not respond to this part of the spacetime. However, we refer the readers to ref. \cite{Bic_aacute_k_2002} for a discussion on some interesting subtleties involved in defining such spacetimes. 

\section{Massless scalar field}
\label{sec: field}

The massless scalar field equation in the above spacetime admits the mode decomposition:
\begin{equation}
    \label{eq:fieldin}
    \Psi_{k m q} = N_{k m q} e^{-i \omega t} e^{i k z} e^{i m \phi} \psi_{m q}(\rho_\pm),
\end{equation}
with $\omega^2 \equiv q^2 +k^2$ and $N_{k m q}$ being a normalisation constant.
Inside the cylinder, $\rho_-<R_1$, the radial equation governing $\psi_{m q}(\rho_-)$ reads
\begin{equation}
    \Bigl(q^2-\frac{m^2}{\rho_-^2}\Bigr)\psi_{m q}(\rho_-)+\frac{\psi_{m q}'(\rho_-)}{\rho_-}+\psi_{m q}''(\rho_-)=0\,,
\end{equation}
which admits the general solution
\begin{equation}
    \psi_{m q}(\rho_-) = a_1 J_{|m|}(q \rho_-)+a_2 Y_{|m|}(q\rho_-),
\end{equation}
where $J_m$ and $Y_m$ are the Bessel functions of the first and second kind, respectively. To impose regularity at $\rho_-=0$, we set $a_2=0$, and to impose periodicity in $\phi$, we have $m\in \mathbb Z$. We can take $a_1=1$ by absorbing it into the normalisation constant $N_{k m q}$.

Meanwhile, the radial equation outside the cylinder reads:
\begin{equation}
    \Bigl(q^2-\frac{c^2 m^2}{\rho_+^2}\Bigr)\psi_{ mq}(\rho_+)+\frac{\psi_{mq}'(\rho_+)}{\rho_+} +\psi_{m q}''(\rho_+)=0\,.
\end{equation}
This admits the general solution
\begin{equation}
    \psi_{mq}(\rho_+) = c_1 J_{|cm|}(q \rho_+)+c_2 Y_{|cm|}(q\rho_+)\,.
\end{equation}
The arbitrary constants $c_1$ and $c_2$ are determined by the continuity of $\psi_{m q}$ and 
%$e^{\mu}_{\rho} \partial_{\mu} \psi_{m q}$ \tcr{\bf we cannot say this -- better `continuity of the normal derivative in the orthonormal frame?} 
its derivative on the cylinder:
\begin{align}
    \psi_{m q}(\rho_-=R_1)&=\psi_{m q}(\rho_+=c R_1)\,,\\
    \partial_{\rho_-}\psi_{m q}(\rho_-=R_1)&=\partial_{\rho_+}\psi_{m q}(\rho_+=c R_1)\,,
\end{align}
which gives
\ba
    c_1(q)&=&\frac{1}{2}\pi q c R_1\Bigl(J_{|m|}(q R_1)Y_{|cm|-1}(q c R_1)\nonumber\\
    &&\quad -J_{|m|-1}(q R_1)Y_{|cm|}(q c R_1)\Bigr)\,,\nonumber\\
    c_2(q)&=&\frac{1}{2}\pi q c R_1\Bigl(J_{|m|-1}(q R_1)J_{|cm|}(q c R_1)\nonumber\\
    &&\quad -J_{|m|}(q R_1)J_{|cm|-1}(q c R_1)\Bigr)\,.
\ea
%\begin{widetext}
%\begin{align}
%    c_1(q)&=\frac{1}{2}\pi q R_1\big(J_{|m|}(q %R_1)Y_{|cm|-1}(q c R_1)-J_{|m|-1}(q R_1)Y_{|cm|}%(q c R_1)\big)\\
%    c_2(q)&=\frac{1}{2}\pi q R_1\big(J_{|m|-1}(q %R_1)J_{|cm|}(q c R_1)-J_{|m|}(q R_1)J_{|cm|-1}(q %c R_1)\big)\,.
%\end{align}
%\end{widetext}
To impose the Dirichlet boundary condition at $R_2$, we restrict  the `radial quantum number'  $q$ to the discrete set such that $\psi_{mq}(\rho_+= c R_2)=0$. 

Finally, the $N_{kmq}$'s are chosen such that the solutions are normalised with respect to the Klein--Gordon (KG) inner product. This gives
\begin{equation}
    N_{k mq} = \frac{1}{2\pi\sqrt{2\sqrt{k^2+q^2}} ||\psi_{mq}||}\,,
\end{equation}
with
\begin{align}
\label{eq: norm}
        ||\psi_{mq}||^2&=\int_0^{R_1} [J_{|m|}(q \rho)]^2 \rho d\rho \nonumber\\&+\int_{c R_1}^{c R_2} [c_1J_{|cm|}(q \rho)+c_2Y_{|cm|}(q\rho)]^2(\rho/c)\, d\rho\,.
\end{align}
It can be checked that the solutions defined above are orthogonal with respect to the KG inner product. 

We can then proceed with canonical quantisation of the field by defining the field operator as
\be
    \hat\psi(x) =\sum_{m, q}\int_{-\infty}^{\infty}\!\!\!\!dk\Bigl(\hat{a}_{kmq}\Psi_{kmq}(x)+\hat{a}^{\dagger}_{kmq}\Psi^{\dagger}_{kmq}(x)\Bigr)\,,
\ee
where $\hat a_{kmq}$ and $\hat a^{\dagger}_{kmq}$ are annihilation and creation operators respectively satisfying the usual commutation relations. We will let $\ket{0}_{F}$ denote the vacuum state of the field satisfying $\hat a_{kmq}\ket{0}_{F}=0$ for all $\hat a_{kmq}$.

\section{Detector response}
\label{sec: UDW}

In this section, we will derive the expression for the response function, $\mathcal{F}$, of a UDW detector placed inside the first cylinder. The detector itself is a simple two-level, point-like quantum system \cite{Unruh:1976db,DeWitt:1980hx}. We will denote the ground and excited states of the detector using $\ket{0}_D$ and $\ket{1}_D$ respectively, and let $\Omega$ denote the energy difference between these two states. In general, the detector can have any arbitrary trajectory in the spacetime of interest. For our purposes   it is enough to consider simple stationary trajectories where the detector stays at a fixed spatial position. In terms of the proper time $\tau$ of the detector, these trajectories are given by:
\begin{align}
    \sx(\tau) &\equiv (t(\tau), \rho_-(\tau), z(\tau), \phi(\tau)) 
    \nonumber\\& = (\tau, \rho_d, z_d, \phi_d),
\end{align}
for $\rho_d\in [0, R_1]$, $z_d\in(-\infty,\infty)$ and $\phi_d\in(0,2\pi]$.
%\tcr{\bf what might be interesting is to consider uniformly accelerated detector along the $z$-axis. Provided the horizon is hiding the cylindrical shell -- what do we see?}

As the detector travels along its trajectory, it interacts locally with the background quantum field. This interaction is modelled by the UDW interaction Hamiltonian which, in the interaction picture, reads:
\begin{equation}
     \hat H_I(\tau) = \lambda\chi(\tau)\,(e^{-i\Omega\tau}\ket{1}_D\bra{0}_D+e^{i\Omega\tau}\ket{0}_D\bra{1}_D)\,\otimes\hat\psi(\sx(\tau))\,.
\end{equation}
In this expression, $\lambda$ is a dimensionless coupling constant and $\chi(\tau)$ is a continuous, compact function describing the relative coupling strength in time. It is aptly called the ``switching function'' of the detector \cite{Satz:2006kb}. Following \cite{cong:2020crf} we will use the following switching function: 
\begin{align}
    \chi(\tau) &= \begin{cases}
      \cos^4(\eta \tau), &-\frac{\pi}{2 \eta}\leq \tau\leq\frac{\pi}{2 \eta}\\
       0, &\text{otherwise\,.}
   \end{cases}\label{eq:compact}
\end{align}
This switching function goes continuously to zero at $\tau = \pm \frac{\pi}{2 \eta}$, i.e., the detector and the field   interact  only for a finite time interval of $\Delta \tau = \pi/\eta$.

If the detector and the field were initialised in the state $\ket{0}_D\ket{0}_F$, there might be a non-zero probability of finding the detector in the state $\ket{1}_D$ at the end of the interaction. This probability can be computed using perturbation theory and is given by \cite{birrell_davies_1982,Pozas2015}
\begin{equation}
\label{eq:P}
\begin{split}
   P = \lambda^2&\int_{-\infty}^{\infty}  d \tau_1\,\int_{-\infty}^{\infty}d \tau_2 \chi  (\tau_1)\chi(\tau_2) e^{-i\Omega(\tau_2-\tau_1)}\\&\times W(\sx(\tau_1),\sx(\tau_2))
    \end{split}
\end{equation}
to second order in $\lambda$. The term $W(\sx(\tau_1),\sx(\tau_2))$ is the Wightman function of the field evaluated along the detector trajectory, $W(\sx(\tau_1),\sx(\tau_2)) := 
\bra{0}{\hat\psi(\sx(\tau_2))\hat\psi(\sx(\tau_1))}\ket{0}$. It can be expanded in terms of the modes given in the previous section as follows:
\begin{equation}\label{Wight}
     W(\sx(\tau_1),\sx(\tau_2))=\sum_{m,q}\int_{-\infty}^{\infty}dk \, \Psi^{\dagger}_{kmq}(\sx(\tau_1))\Psi_{kmq}(\sx(\tau_2))\,.
\end{equation}

Substituting this mode expansion into Eq.~\eqref{eq:P} gives, 
\ba \label{eq: resp}
    \mathcal{F} &=&\int_{-\infty}^{\infty}\int_{-\infty}^{\infty} dt\, dt' \chi(t)\chi(t')e^{-i \Omega (t-t')} \nonumber\\
    &&\quad \times \sum_{m,q}\int_{-\infty}^{\infty}dk N^2_{kmq}|J_{|m|}(q\,\rho_d)|^2 e^{-i \omega (t-t')}
   \nonumber\\ 
   &=& \sum_{m q} \frac{|J_{|m|}(q \rho_d)|^2}{||\psi_{mq}||^2 4\pi}\int\frac{|\hat\chi(\Omega+\sqrt{k^2+q^2})|^2}{\sqrt{k^2+q^2}}dk\quad
\ea
for the \textit{response function} $\mathcal{F}\equiv P/\lambda^2$. Here, the function $\hat \chi$ is the Fourier transform of the switching function,
\begin{equation}
    \hat\chi(y)=\frac{1}{\sqrt{2\pi}}\int_{-\infty}^{\infty}d\tau\chi(\tau)e^{-iy\tau}\,.
\end{equation}
Note that the response of the detector depends, as expected, only on radial coordinate of the detector, and not on its $\phi$ and $z$ coordinates due to the symmetry of the set-up. In addition, the response does have an implicit dependence on the conicity parameter $c$ coming from the normalisation term $||\psi_{mq}||^2$, see~\eqref{eq: norm}. The results of the next section will illustrate this dependence. 

\section{Results}
\label{sec: results}

Having introduced the set-up and derived the response of the detector in the previous sections, we are now ready to look at how the response  
$\mathcal{F}_c$  depends on the conicity $c$. The response of a detector placed in a spacetime with no conical deficit (corresponding to $c=1$) will be denoted as $\mathcal{F}_{c=1}$.

\begin{figure}
    \centering
    \includegraphics[scale=0.5]{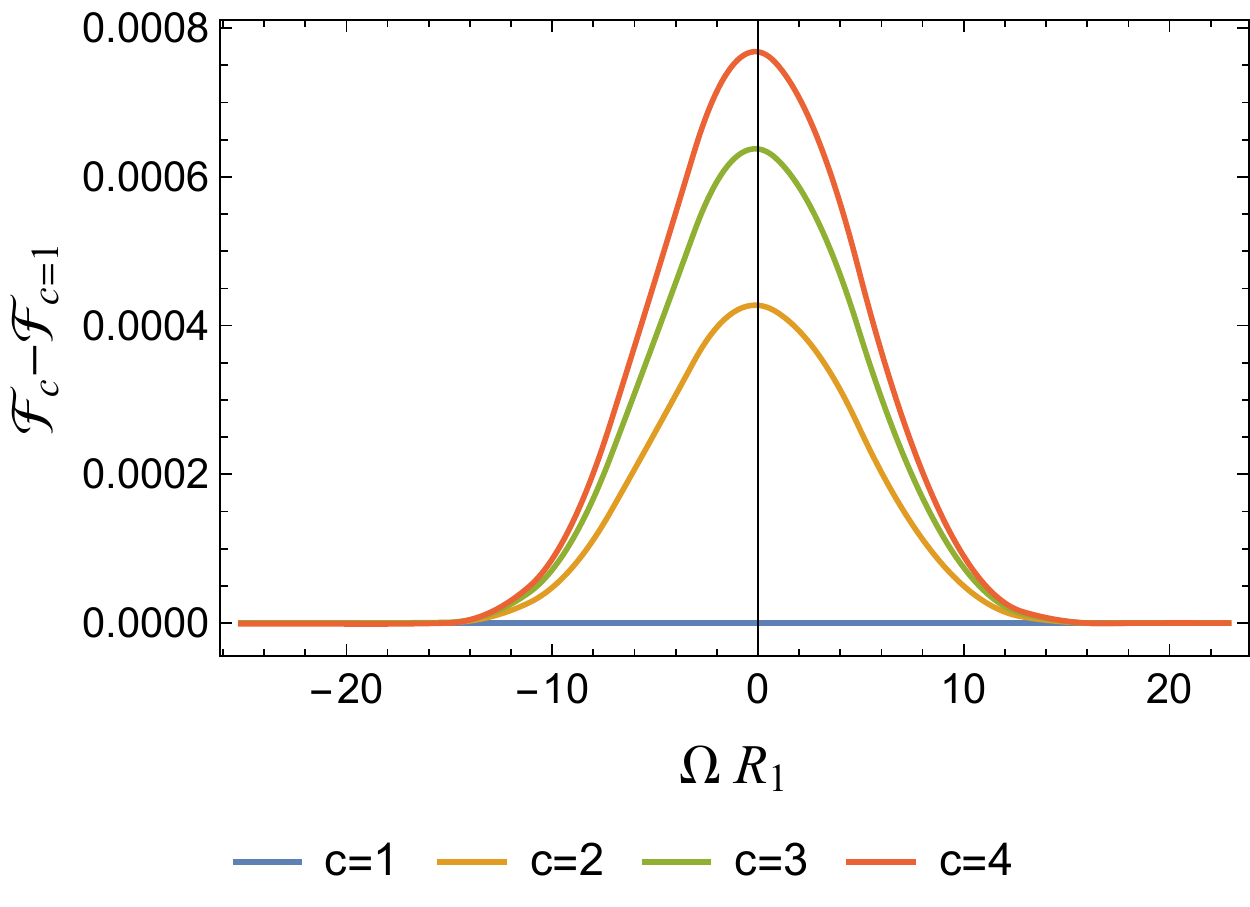}
    \includegraphics[scale=0.5]{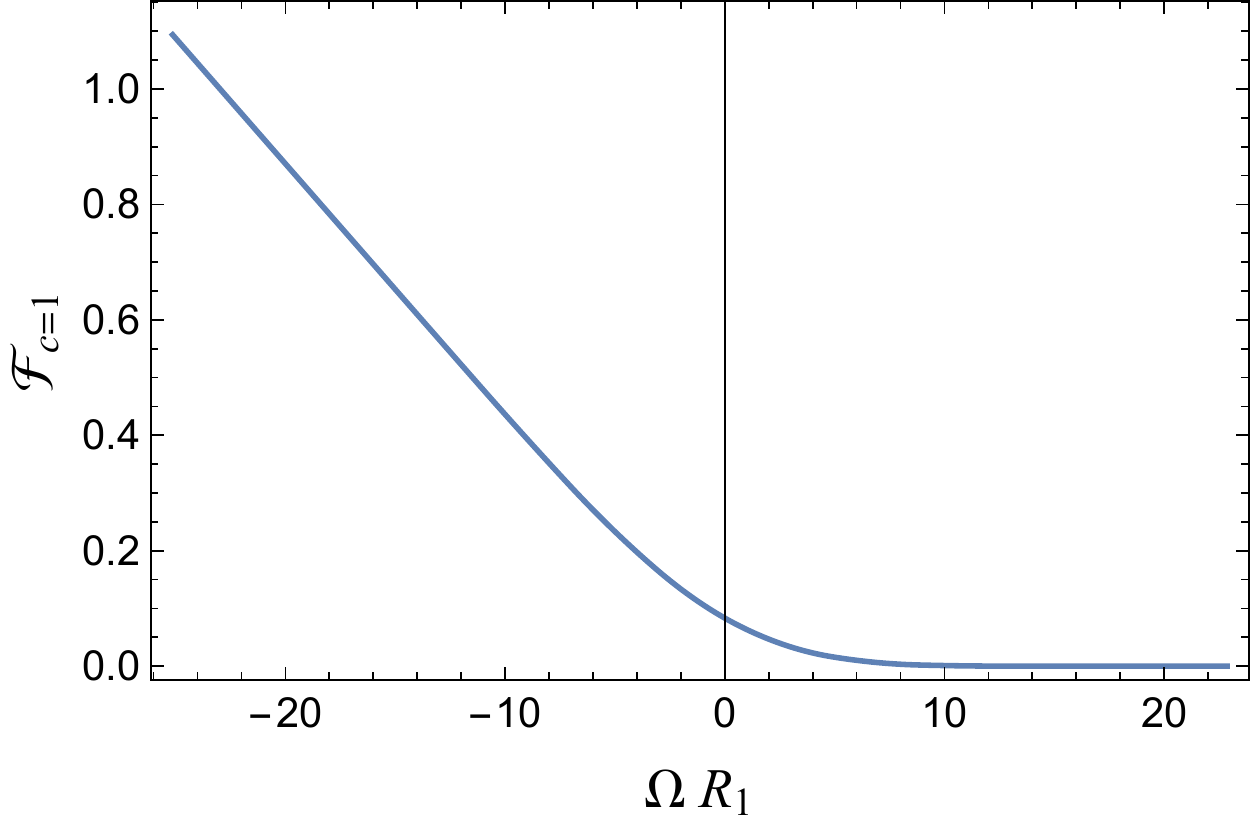}
    \caption{Response against detector energy gap. \textbf{Top:} This figure shows how the difference $\mathcal{F}_c-\mathcal{F}_{c=1}$ varies with $c$ and the detector energy gap $\Omega$. The difference appears to be symmetric in $\Omega$, and peaks at $\Omega=0$. The magnitude of the difference also increases with $c$. \textbf{Below:} This figure shows the general shape of $\mathcal{F}$ as a function of $\Omega$. The value of $c$ used here was $c=1$; the corresponding curves for the other $c$ values in the top plot will simply overlap with the existing curve due to scale of the figure. The other parameters used here are $R_1 = 1$, $R_2=5$, $\rho_d=0$ and $\Delta \tau=1$.}
    \label{fig: respvgap}
\end{figure}

\begin{figure}
    \centering
    \includegraphics[scale=0.5]{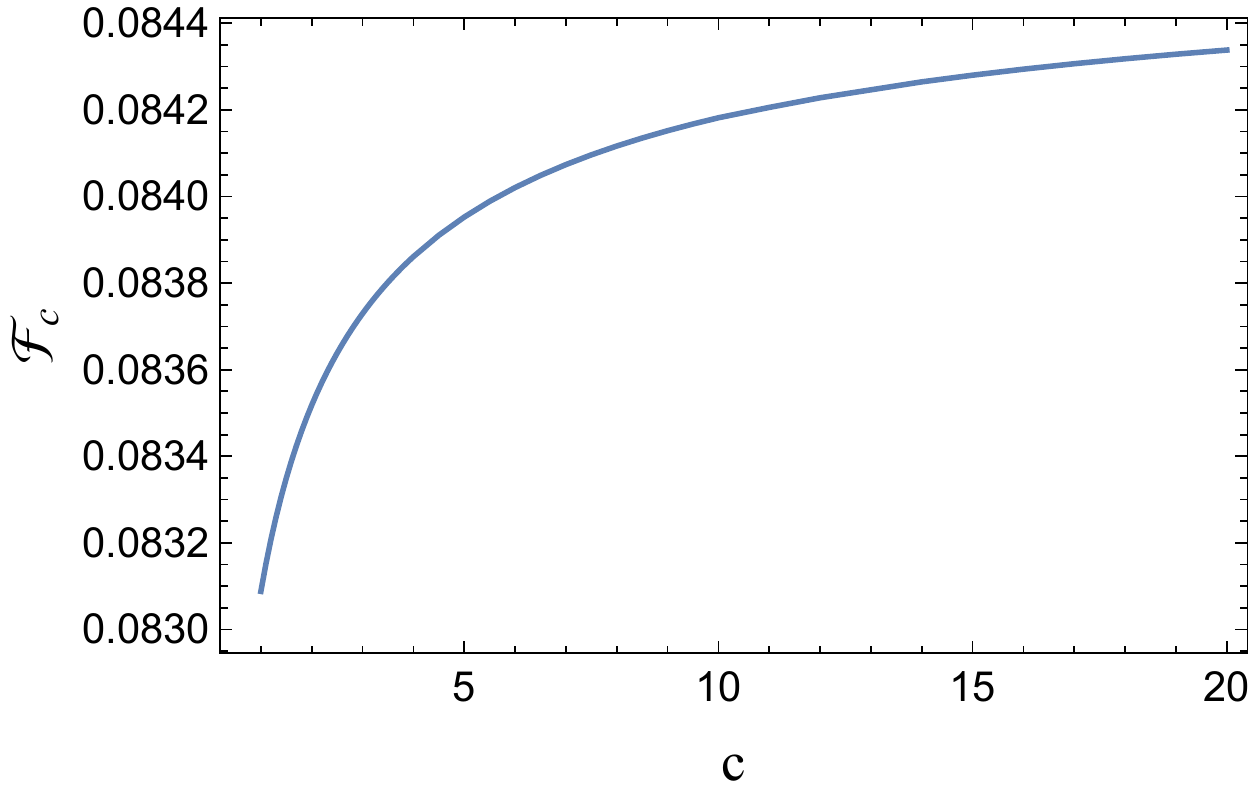}
    \caption{Dependence of response on $c$. This figure gives a plot of $\mathcal{F}_c$ against $c$ for $R_1 = 1$, $R_2=5$, $\Omega=0$, $\rho_d=0$ and $\Delta \tau=1$.}
    \label{fig: respvc}
\end{figure}

The top figure in Fig. \ref{fig: respvgap} shows a plot of $\mathcal{F}_c-\mathcal{F}_{c=1}$ against $\Omega$ for $c\in\{1,2,3,4\}$. From this we see that the difference is indeed non-zero, peaks around $\Omega=0$, and increases as $c$ increases. In this plot, we have set $\Delta\tau = R_1=1$, so that the detectors are switched on only for a short duration, during which no signal could have travelled from the detector to the shell and back to convey information about $c$. Intuitively, the dependence of the response on $c$ conveys the fact that the local vacuum fluctuations around the detector carry non-local information about the spacetime. The dependence of $\mathcal{F}$ on $c$ can be seen more clearly in Fig. \ref{fig: respvc}, which gives a plot of $\mathcal{F}_c$ against $c$ for $\Omega=0$. This graph shows a logarithmic increase in the response of the detector as $c$ increases. %\rbm{This is a bit strange -- if $\Omega=0$ the detector can't get excited.  This is
%some process where the detector begin in one of its degenerate energy levels, with the other one becoming (partly) filled after the interaction.
%}

\begin{figure}
    \centering
    \includegraphics[scale=0.5]{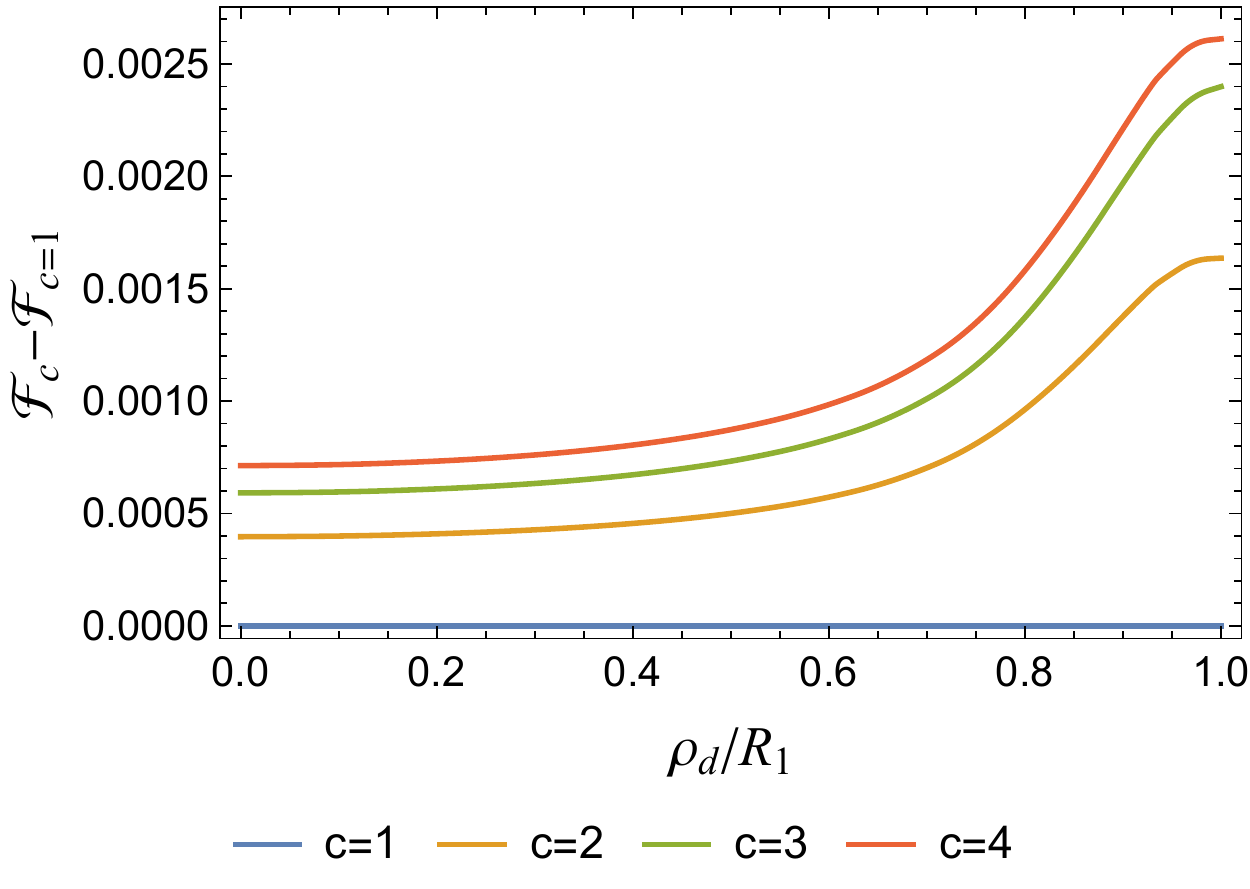}
    \caption{Dependence of response on detector radial distance. The figure here shows that the difference $\mathcal{F}_c-\mathcal{F}_{c=1}$ increases as $\rho_d$ increases i.e., the detector gets closer to the cylindrical shell. The parameters used here are $R_1 = 1$, $R_2=5$ and $\Omega=2$.}
    \label{fig: respvrd}
\end{figure}

The results shown in Fig. \ref{fig: respvgap} were for $\rho_d=0$, with the detector   placed on the axis of symmetry. This greatly reduces the computational effort since $J_m(0) = \delta_{m,0}$ and only the $m=0$ term in Eq.~\eqref{eq: resp} contributes. Fig. \ref{fig: respvrd} shows that the difference in response increases as the detector is moved closer to the cylindrical shell.

\begin{figure}
    \centering
    \includegraphics[scale=0.5]{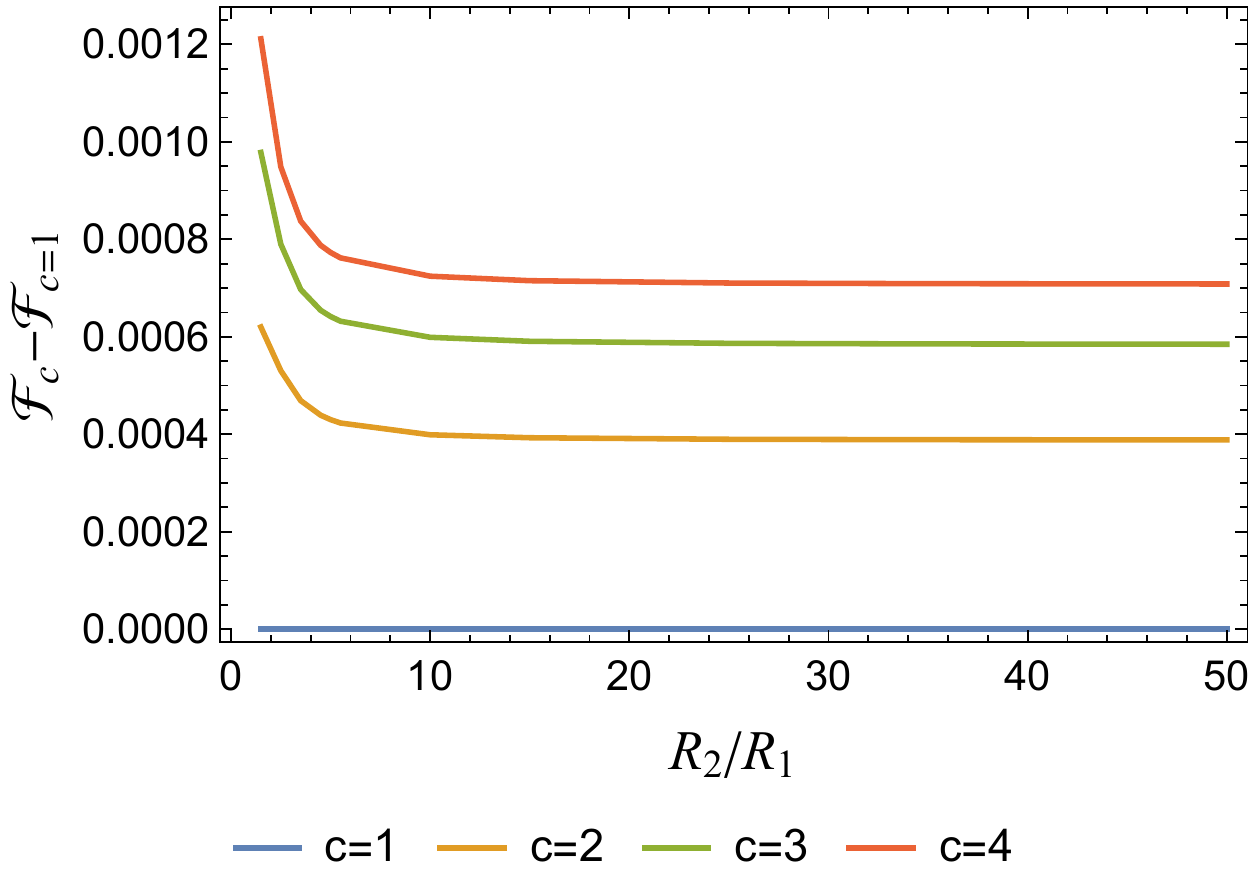}
    \includegraphics[scale=0.5]{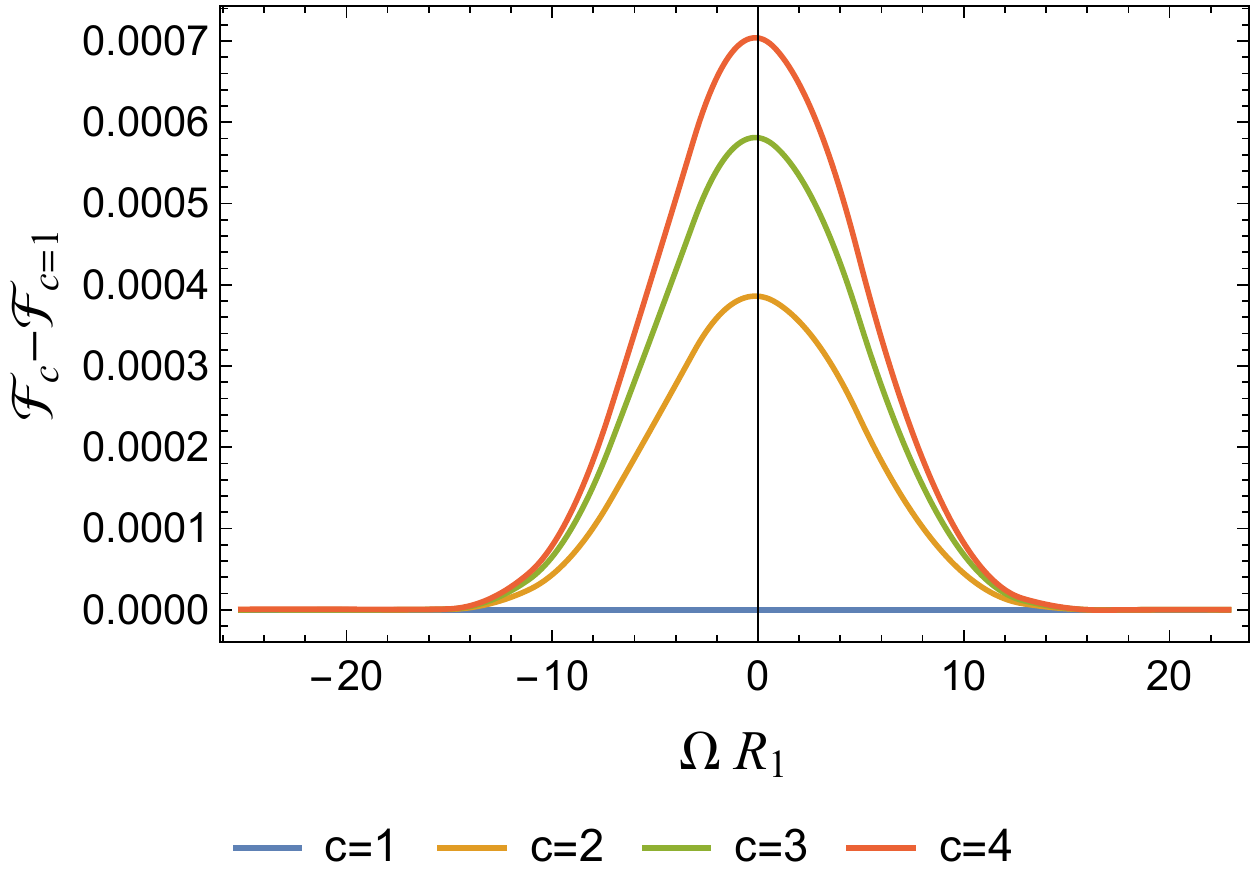}
    \caption{Dependence of response on $R_2$. \textbf{Top:} The figure shows a plot of $\mathcal{F}_c-\mathcal{F}_{c=1}$ against $R_2$, the position of the Dirichlet boundary, for various $c$ values. The difference asymptotes to some non-zero value at large $R_2$. \textbf{Below:} Plot of $\mathcal{F}_c-\mathcal{F}_{c=1}$ against $\Omega$ (cp Fig. \ref{fig: respvgap}) for $R_2=50$. The other parameters used here are $R_1=1$ and $\rho_d=0$.}
    \label{fig: respvR2}
\end{figure}

Recall that we have discretised the integral over $q$ into a discrete sum by introducing a Dirichlet boundary condition at $\rho_+ = c R_2$. Fig. \ref{fig: respvR2} shows what happens as we push this boundary outwards. From the top figure, we see that the difference in response asymptotes to some finite value as $R_2$ increases. Meanwhile, plotting the difference in response against $\Omega$ at $R_2=50$ displays the same trends as Fig. \ref{fig: respvgap} but with 
slightly decreased magnitude. However, based on the top figure, we can expect non-zero differences even when the Dirichlet boundary is pushed out towards infinity.

\section{Conclusion}
\label{sec: conclusion}

We have shown that a conical deficit exerts a detectable influence on the response of a UDW detector even if that detector is in a flat spacetime region without access to the region of spacetime where the deficit is manifest. An observer restricted to the same region with access only to classical measuring devices would not be able to detect the presence of the deficit outside the cylinder. However the UDW detector can discern the presence of the deficit.   

This situation is similar to that for a detector located inside a spherical shell:  it can read out information about the non-local structure of spacetime even when switched on for scales much shorter than the characteristic scale of the non-locality \cite{Ng:2016hzn}. As with the spherical shell, we find the sensitivity to the deficit is strongest at vanishing energy gap for a detector located on the axis of the cylinder, and increases as the detector is located further from the axis. 

A number of future studies merit consideration. Can further information be extracted via this process so that the  cylinder's density profile or rotation can be determined?  Can one obtain similar results for  more realistic detector models interacting with electromagnetic fields?   

Finally, we note that one may also consider placing the UDW detector outside the first cylinder. In this case, the
$R_1 \to 0$ limit will correspond to a cosmic string placed
inside a reflective concentric cylinder. We can see this
by noting that $c_2(q) \to 0$ in this limit and the solution $\psi_{mq}(\rho_+)$ in Eq. (8) reduces to that in the cosmic string
spacetime after normalisation. The model can then be used to compare, for example, the difference between the
cases when the string is modelled as a Dirac delta source
or as a finite cylinder. Work on these areas is in progress.
 
 \section*{Acknowledgments}
 \label{sc:acknowledgements}
 This work was supported in part by the Natural Sciences and Engineering Research Council of Canada,  the Perimeter Institute, and Asian Office of Aerospace Research and Development Grant FA2386-19-1-4077. J.B. thanks the kind hospitality of the Perimeter Institute, Waterloo, where this work started. J.B. also acknowledges the support from the Grant Agency of the Czech Republic, Grant No. GA\u CR 21/11268S. Research at Perimeter Institute is supported in part by the Government of Canada through the Department of Innovation, Science and Economic Development Canada and by the Province of Ontario through the Ministry of Colleges and Universities. 
Perimeter Institute and the University of Waterloo are situated on the Haldimand Tract, land that was promised to the Haudenosaunee of the Six Nations of the Grand River, and is within the territory of the Neutral, Anishnawbe, and Haudenosaunee peoples.
 
 \appendix
 \begin{widetext}
 \section{Stress-Energy of Cylinder}
 \label{appendix}
 
  In this Appendix we briefly review the stress-energy tensor of the cylindrical shell following the discussion in \cite{Bic_aacute_k_2002}. The general line element of a spacetime admitting cylindrical symmetry  has the form,
 \begin{equation}
     ds^2 = -e^{2U(\tilde r)} d\tilde t^2 +e^{-2U(\tilde r)}[A(\tilde r)^2d\tilde r^2+B(\tilde r)^2d\varphi^2+C(\tilde r)^2 d\tilde z^2]\,.
 \end{equation}
By substituting this ansatz into the vacuum Einstein's field equations,  after appropriate rescaling and translation of the coordinates, the general solution has three arbitrary constants, $r_0, M$ and $c$  \cite{Bicak:2004fw}:
 \begin{equation}
     ds^2 = -\big(\frac{r}{r_0}\big)^{2M} d\tau^2 +\big(\frac{r}{r_0}\big)^{2M(M-1)}(d\zeta^2+dr^2)+r^2 \big(\frac{r}{r_0}\big)^{-2M}d\varphi^2/c^2,
 \end{equation}
 with the coordinate ranges $\tau\in \mathbb{R}$, $r\in \mathbb{R}^+$, $\zeta\in\mathbb{R}$ and $\varphi\in [0,2\pi)$. However, we can get rid of $r_0$ by another coordinate rescaling 
 %$\rho \equiv r/r_0,\, t \equiv \tau/r_0,\,z\equiv\zeta/r_0$. 
 $$(\tau,\zeta,r,\varphi) \rightarrow (r_0^{\frac{M}{M^2-M+1}}t,r_0^{\frac{M(M-1)}{M^2-M+1}}z, r_0^{\frac{M(M-1)}{M^2-M+1}}\rho,\varphi)\,,
 $$
 so that the final metric reads:
 \begin{equation}
 \label{eq:LC}
     ds^2 = -\rho^{2M} dt^2+\rho^{-2M}\bigg[\rho^{2M^2}(dz^2+d\rho^2)+\rho^2 d\varphi^2/c^2\bigg]\,.
 \end{equation}
We are thus left with only 2 physical parameters $M$ and $c$ for the spacetime. This is the common form of the Levi-Civita metric appearing in the literature.
 
We now move on to discuss the cylindrical shell that can give rise to such a metric. A general cylindrical shell spacetime consists of two regions, each described by the metric in Eq.~\eqref{eq:LC} with specific parameter choices: an ``inside'' region which is regular along the $z$-axis with $c_-=1$ and $M_-=0$ running from $0\leq \rho_-\leq R_-$, and an ``outside'' region with arbitrary $c_+$ and $M_+$ running from $\tilde R_1 \leq \rho_+ \leq \tilde R_2$. These two regions are joined at the shell surface using the Israel formalism, which sets $R_1 = \tilde R_1^{1-M_+}/c$, as well as the non-vanishing components of the stress-tensor of the shell as:
 \begin{align}
     8\pi S_{tt} &= 1/R_1-\tilde R_1^{M_+-M_+^2-1}(1-M_+)^2\,,\nonumber\\
     8\pi S_{zz} &= \tilde R_1^{M_+-M_+^2-1}-1/R_1\,,\nonumber\\
     8\pi S_{\varphi\varphi} &= \tilde R_1^{M_+-M_+^2-1}M_+^2\,.
 \end{align}
 We can define the mass per unit coordinate length of the cylinder as $\mu = 2\pi R_1 S_{tt}$, which in terms of the spacetime parameters, reads
 \begin{equation}
     \mu = \frac{1}{4}\bigg(1-\frac{1}{c_+}\frac{(1-M_+)^2}{\tilde R_1^{M_+^2}}\bigg)\,.
 \end{equation}
 As we were only interested in studying the sensitivity of a UDW to the conical deficit outside the cylinder, we have set $M_+ = 0$ in the main text to avoid any effects on the UDW due to non-trivial curvature outside the cylinder. However, we see here that even in this case, the shell has a non-zero mass per unit length, given by \eqref{mu} in the main text.
 %of $  = \frac{1}{4}(1-\frac{1}{c_+})$.
 
\end{widetext}
\bibliography{CylinderUDW11}

\end{document}